\documentclass{PoS}
\usepackage[dvips]{graphicx}
\usepackage{psfrag}
\usepackage{amsmath,amssymb}

\PoS{PoS(LAT2005)248}

\title{Vortex Proliferation and the Dual Superconductor Scenario for
Confinement: \\ The 3D Compact U(1) Lattice Higgs Model}

\ShortTitle{Vortex Proliferation and the Dual Superconductor Scenario for
Confinement}

\author{\speaker{Sandro Wenzel}\thanks{S.W. acknowlegdes financial
        support from the Studienstiftung des deutschen Volkes. This work
        is further partially supported by DFG Grant No. JA 483/17-3.},
        Elmar Bittner, Wolfhard Janke, Adriaan
        M. J. Schakel\thanks{Present address: Institut f\"ur Theoretische
        Physik, Freie Universit\"at Berlin, Arnimallee 14, 14195 Berlin,
        Germany.}, \hbox{and Arwed Schiller}\\ Institut f\"ur
        Theoretische Physik, Universit\"at Leipzig,\\ Augustusplatz
        10/11, D-04107 Leipzig, Germany\\ E-mail:
        \email{[wenzel|bittner|janke|schakel|schiller]@itp.uni-leipzig.de}}

\abstract{It is argued that the phase diagram of the 3D Compact U(1)
Lattice Higgs Model is more refined than generally thought. The confined
and Higgs phases are separated by a well-defined phase boundary, marked
by proliferating vortices.  It is shown that the confinement mechanism
at work is precisely the dual superconductor scenario.}

\FullConference{XXIIIrd International Symposium on Lattice Field Theory\\
                 25-30 July 2005\\
                 Trinity College, Dublin, Ireland}

\newcommand{\dd}{\mathrm{d}}

\begin{document}

\section{Introduction}

Since a confining state is a very special state of matter, it would be
desirable for it to be clearly distinguishable from any other state the
system can assume.  Ideally, the confining state would be separated from
the other states by a phase transition with an order parameter signaling
the fundamental change of the ground state.  However, by studying the
three-dimensional (3D) Abelian Higgs model with compact gauge field,
Fradkin and Shenker \cite{FradkinShenker} provided a counterexample.
For a Higgs field carrying one unit charge $q=1$, they showed that in
the London limit, where the amplitude of the Higgs field is kept fixed,
it is always possible to move from the Higgs region into the confined
region without encountering singularities in local gauge-invariant
observables.  As for the liquid-vapor transition, this is commonly
interpreted as implying that the two ground states do not constitute
distinct phases.  This can be further supported by symmetry
considerations \cite{KovnerRosenstein}.  The relevant \textit{global}
symmetry for the model with a Higgs field carrying charge $q$ is the
cyclic group Z$_q$ of $q$ elements.  For $q=2$, this implies
the possibility that the confined and Higgs phases are separated by a
continuous phase transition belonging to the 3D Ising universality
class.  This is indeed known to be the case
\cite{FradkinShenker,Bhanot:1981ug}.  For $q=1$ on the other hand, this
simply excludes a continuous phase transition as the group
$Z_1$, consisting of only the unit element, cannot be spontaneously
broken.  We are thus left with the rather unsatisfying situation of
having a confined state which is not at all that special, being
analytically connected to a different ground state with different,
unrelated physical properties, viz\ the Higgs phase.  To address this
issue, we numerically investigate the $q=1$ model with fluctuating Higgs
amplitude.  Similar studies were already carried out as early as in 1985
\cite{Munehisa:1985rb,Obodi:1985uu} on smaller lattices, and more
recently in Refs.~\cite{Kajantie:1997vc,Chernodub:2004yv} on larger
ones.  A first report on our findings recently appeared in
Ref.~\cite{Leipzighiggs}.

\section{Monte Carlo Simulations}
The compact Abelian lattice Higgs model is defined by the 3D Euclidean
lattice action $S = S_g + S_\phi$.  The gauge part is the standard
action for a compact Abelian gauge field
\begin{equation}
\label{Sg} 
  S_g = \beta\sum_{x,\mu<\nu} \left[ 1-\cos \theta_{\mu \nu} (x) \right],
\end{equation} 
where $\beta$ is the inverse gauge coupling.  The sum extends over all
lattice sites $x$ and lattice directions $\mu$, and $\theta_{\mu
\nu}(x)$ denotes the plaquette variable $\theta_{\mu \nu} (x) =
\Delta_\mu \theta_{\nu} (x) - \Delta_\nu\theta_{\mu} (x)$, with the
lattice derivative $\Delta_\nu \theta_{\mu} (x) \equiv \theta_{\mu}
(x+\nu) - \theta_{\mu} (x)$ and the compact link variable $\theta_{\mu}
(x) \in[-\pi,\pi)$.  The matter part of the action consists of a
$|\phi|^4$ theory minimally coupled to the gauge field
\begin{equation}
 S_\phi \!\! = \!\! -\kappa\sum_{x,\mu}\rho(x) \rho(x+\mu) \cos 
 \left[\Delta_\mu\varphi(x) - \theta_{\mu}(x)\right] 
  + \sum_x \left\{\rho^2(x) + \lambda 
   \left[\rho^2(x)-1\right]^2 \right\},
\end{equation}
where the complex Higgs field is represented by its amplitude and phase
$\phi(x) = \rho(x) \mathrm{e}^{\mathrm{i} \varphi(x)}$, with $\varphi(x)
\in[-\pi,\pi)$.  The parameter $\kappa$ is the hopping parameter and
$\lambda$ the Higgs self-coupling, which together with the inverse
coupling constant $\beta$ constitute the parameters of the theory. The
model is put on a cubic lattice of linear size $L$ with periodic
boundary conditions.  The pure $|\phi|^4$ theory with fluctuating
amplitude, obtained by taking the gauge coupling to zero, i.e., by
letting $\beta \to \infty$, was recently investigated by means of Monte
Carlo simulations in Ref.~\cite{bittner}.

We monitor observables that probe the gauge and matter parts separately.
For the gauge part we consider the monopole density $M$ \cite{degrand}
and the Polyakov loop, as was done earlier in Ref.~\cite{schiller1},
where the London limit of the model was considered.  For the matter part
we consider the Higgs amplitude squared $\rho^2\equiv(1/L^3)\sum_x
\rho^2(x)$.  In addition, we monitor the plaquette action (\ref{Sg})
(divided by $3L^3$) and the link observable $C = - (1/3 L^3)
\sum_{x,\mu} \cos \left[\Delta_\mu\varphi(x) - q \theta_{\mu}(x)
\right]$.  Both Metropolis and heat-bath methods are used to generate
Monte Carlo updates.  Since these local updates become inefficient at
first-order phase transitions, we implement the multicanonical method
\cite{berg1} and reweighting techniques \cite{ferrenberg} to access
these regions of phase space.  The simulations are carried out at
\textit{fixed} inverse gauge coupling $\beta$ on cubic lattices varying
in size from $6^3$ to $32^3$, in extreme cases to $42^3$.
Thermalization of production runs typically take $4 \times 10^4$ sweeps
of the lattice, while about $10^6$ sweeps are used to collect data, with
measurements taken after each sweep of the lattice.  Because of its
pronounced peaks, we use the maxima of the link susceptibility
$\chi_C=L^3 \left( \langle C^2\rangle - \langle C \rangle^2 \right)$
together with histograms, rescaled to equal height, to trace out the
phase boundary.  Statistical errors are estimated by means of jackknife
binning.  For a detailed description of the algorithms and their
implementation, the reader is referred to Ref.~\cite{wenzel}.

\section{Phase Diagram}
\begin{figure}
\begin{minipage}{0.6\textwidth}
\includegraphics[width=0.95\textwidth]{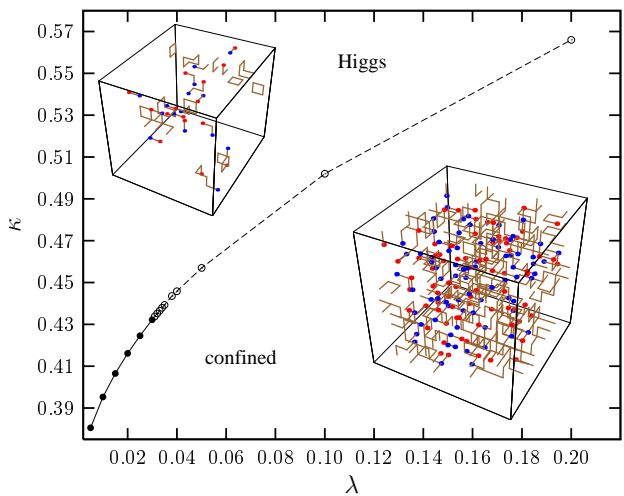}
\end{minipage}
\begin{minipage}{0.38\textwidth}
\vspace*{0.5cm}
\caption{\label{fig:phasediagram} $\kappa$-$\lambda$ phase diagram at
    $\beta=1.1$ in the infinite-volume limit. Solid dots mark the
    first-order phase transition line ending
    at $0.030<\lambda_\mathrm{c}<0.032$.  Open
    dots for $\lambda > \lambda_c$ mark the location of the Kert\'esz
    line approaching $\kappa=0.717(2)$ in the London limit
    $\lambda\to\infty$. The insets show snapshots of typical monopole
    and vortex configurations in both phases, with red dots denoting
    monopoles and blue dots denoting antimonopoles, each
    monopole-antimonopole pair being connected by a vortex.}
\end{minipage}
\end{figure}
Figure~\ref{fig:phasediagram} summarizes our results for $\beta=1.1$.
We identify two phases: a confined and a Higgs phase which below a
critical point $\lambda_\mathrm{c}(\beta)$ are separated by a
first-order phase transition as was already observed in the earlier
Monte Carlo simulations on smaller lattices
\cite{Munehisa:1985rb,Obodi:1985uu}.  For $\beta=1.1$ we estimate the
critical point, where the first-order line ends, to be located in the
interval $0.030 < \lambda_\mathrm{c}(1.1) < 0.032$ in the infinite
volume limit.  The behavior of the average $\langle \rho^2 \rangle$
identifies the phase in the upper left part of the phase diagram as
Higgs phase, where it increases more or less linearly with increasing
$\kappa$, while it takes on a minimum value in the confined phase. The
average plaquette action takes on a finite value in the confined phase
practically independent of $\kappa$ and $\lambda$, while it decreases
with increasing $\kappa$ in the Higgs phase. The identification of the
two phases is consistent with the behavior of the Polyakov loop we
observed.

Of more importance from the perspective of understanding the mechanism
leading to charge confinement is the behavior of the monopole density.
Monopoles and vortices appear due to compactness of the
phase angles $\theta_\mu(x)$ and $\varphi(x)$, respectively.
Using the compact notations of
differential forms on the lattice~\cite{ANO_definition}
the monopole charge and vortex current are defined as follows 
\begin{equation}
\label{eqn:vortex_current}
m = \frac{1}{2 \pi} \dd {[\dd \theta]}_{2\pi} \,, \quad
j = \frac{1}{2 \pi} \left(\dd {[\dd \varphi-\theta]}_{2\pi}+{[\dd \theta]}_{2\pi}\right)\,, 
\end{equation} 
where
${[\cdots]}_{2\pi}$ denotes the integer part modulo $2 \pi$. 
From those constructed quantities we derive the monopole density and 
study the percolation properties.

First of all, within the achieved accuracy, we observed that the
monopole susceptibility $\chi_M$ peaks at the same location as does $\chi_C$.  As
expected, the monopole density is finite in the confined phase, forming
a monopole-antimonopole plasma (see bottom inset in
Fig.~\ref{fig:phasediagram}).  The density is practically independent of
$\kappa$ and $\lambda$.  As $\beta$ increases, the monopole density
decreases.  The monopoles become completely suppressed in the weak gauge
coupling limit $\beta \to \infty$, where the model reduces to the pure
$|\phi|^4$ theory and loses its confining properties.  In the Higgs
phase, the monopole density is vanishing small.  The few monopoles still
present are tightly bound in monopole-antimonopole pairs
\cite{schiller1} (see top inset in Fig.~\ref{fig:phasediagram}). Being
rendered ineffective, the monopoles can no longer confine electric
charges.

\section{Kert\'esz Line}
The nature of the phase boundary is well established in the region
$\lambda < \lambda_\mathrm{c}(\beta)$, where it is a first-order line.
The way the boundary then continues has always been a puzzle.  As seen
already earlier in the London limit \cite{Bhanot:1981ug,schiller1}, for
sufficiently large lattices, the maxima of the susceptibilities do not
show any finite-size scaling.  Moreover, the susceptibility data for the
various observables obtained on different lattice sizes collapse onto
single curves without rescaling, indicating that the infinite-volume
limit is reached.  Since a first-order phase transition can be safely
excluded in this region, the absence of finite-size scaling strongly
suggests the absence of thermodynamic singularities in the
infinite-volume limit, and the boundary, although well defined, is
usually referred to as a mere crossover.

In a previous paper \cite{Leipzighiggs} we argued that the phase
boundary is in fact a \textit{Kert\'esz line}.  Such a line, first
introduced in the context of the Ising model in the presence of an
applied magnetic field \cite{Kertesz}, is now more generally taken as
referring to a well-defined and precisely located phase boundary across
which geometrical objects proliferate, yet thermodynamic quantities
remain nonsingular.  The proliferating objects in the 3D compact Abelian
lattice Higgs model are the vortices defined by the vortex current in Eq.\ \eqref{eqn:vortex_current}.
This picture essentially vindicates the scenario put forward by Einhorn and Savit \cite{einhorn2}
who argued that the transition from the Higgs to the confined phase is
triggered by \textit{proliferating vortices}.  The interpretation of a
deconfinement transition as a Kert\'esz line was earlier proposed in the
context of the SU($2$) Higgs model
\cite{Fortunato:2000ge,Satz:2001zf,Langfeld:2002ic_2,Bertle:2003pj}.
For further examples in the literature, see
Refs.~\cite{Chernodub:1998,Baig:1998}.

\begin{figure}
\begin{minipage}{0.6\textwidth}
\includegraphics[clip,width=0.9\textwidth]{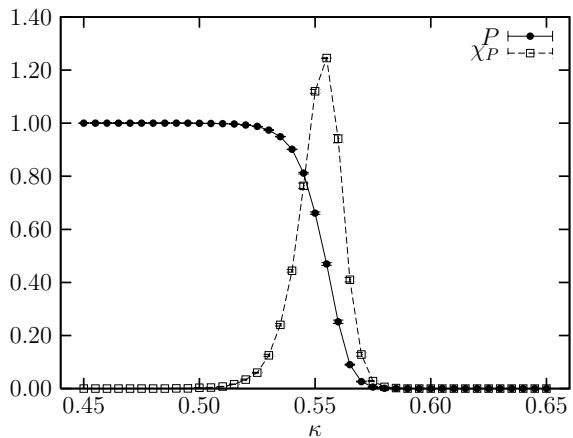}
\end{minipage}\null
\begin{minipage}[t]{0.38\textwidth}
\caption{\label{fig:percolation} Probability $P$ of finding a connected
  vortex network percolating the system of linear size $L=10$ and the
  corresponding susceptibility $\chi_P$ for $\lambda=0.2$ (and
  $\beta=1.1$ as in the rest of the paper).}
\end{minipage}
\end{figure}

We are presently investigating the vortex network directly, using
percolation observables.  As an example, we show in
Fig.~\ref{fig:percolation} a preliminary result for the probability $P$
of finding a connected vortex network percolating the system and the
corresponding susceptibility $\chi_P$.
In the confined phase, for small values of the hopping parameter
$\kappa$, it is always possible to find such a connected vortex network
percolating the system.  In the Higgs phase, on the other hand, this
probability rapidly decreases to zero. The peak of the corresponding
susceptibility is located slightly below the phase boundary as traced
out by the link susceptibility $\chi_C$.  This is similar to what is
found in the pure $|\Phi|^4$ theory \cite{Kajantie:2000cw,bittner3}.
Since vortex percolation cannot be defined unambigously a more
systematic investigation is needed and is currently carried out.

\section{Dual Superconductor Scenario for Confinement}
The confinement mechanism in the 3D compact Abelian lattice Higgs model
is precisely the dual superconductor scenario for confinement
\cite{'tHooft:1977hy}.  In the Higgs phase, the monopoles are tightly
bound together in monopole-antimonopole pairs (see top inset in
Fig.~\ref{fig:phasediagram}).  The magnetic flux emanating from a
monopole is squeezed into a short vortex, carrying one unit $2\pi/q$
($q=1$) of magnetic flux, which terminates at an antimonopole.  The
vortices, which in this phase can also exist as small fluctuating loops
(also seen in the top inset in Fig.~\ref{fig:phasediagram}), have a
finite line tension.  Upon approaching the phase boundary, the vortex
line tension vanishes.  When this happens, the vortices proliferate,
gaining configurational entropy without energy cost, and an infinite
vortex network appears which disorders the Higgs ground state.  At the
same time, the monopoles are no longer bound in tight
monopole-antimonopole pairs but form, as seen in the bottom inset in
Fig.~\ref{fig:phasediagram}, a plasma which exhibits charge confinement.

\section{Conclusions}
We have shown that the phase diagram of the 3D compact Abelian lattice
Higgs model is more refined than hitherto generally thought.  Although
the confined and Higgs phases are analytically connected, a well-defined
phase boundary separating the two phases does exist, marked by
proliferating vortices.  For fixed gauge coupling, the phase boundary is
a line of first-order phase transitions at small Higgs self-coupling,
which ends at a critical point.  The phase boundary then continues as a
Kert\'esz line across which vortices proliferate, yet thermodynamic
quantities and other local gauge-invariant observables remain
nonsingular.  The resulting picture for the confinement mechanism is precisely
the 't Hooft's dual superconductor scenario.

\bibliographystyle{JHEP} \bibliography{literature}

\end{document}